\documentclass[aps,prb,showpacs,twocolumn,amsmath,amssymb,footinbib]{revtex4}
\usepackage{latexsym}
\usepackage{subfigure}
\usepackage{epsfig}
\usepackage{amsfonts}
\usepackage{amssymb}
\usepackage{amsmath}
\usepackage{bm}
\usepackage{braket}
\usepackage{ascmac}
\usepackage{color}
\usepackage{pstricks}%
\usepackage{pst-eps}%
\usepackage{pst-plot}%
\usepackage{pst-node}%
\usepackage{pst-poly}
\newcommand{\Tr}{{\rm Tr}\;}

\renewcommand{\i}{{\rm i}}
\newcommand{\e}{{\rm e}}
\renewcommand{\Im}{{\rm Im}\;}
\renewcommand{\d}{{\rm d}}
\newcommand{\sgn}{{\rm sgn}}
\begin{document}
\title{Detection of topological states in two-dimensional Dirac systems
by the dynamic spin susceptibility}
\author{Masaaki Nakamura$^{1,2}$ and Akiyuki Tokuno$^{3,4}$}
\affiliation{$^1$Institute of Industrial Science, The University of Tokyo,
Meguro-ku, Tokyo, 153-8505, Japan}
\affiliation{$^2$Department of Physics, Ehime University
Bunkyo-cho 2-5, Matsuyama, Ehime 790-8577, Japan}
\affiliation{$^3$Centre de Physique Th\'eorique, Ecole Polytechnique, CNRS,
91128 Palaiseau Cedex, France}
\affiliation{$^4$Coll\`{e}ge de France, 11 place Marcelin Berthelot, 75005
Paris, France}
\date{\today}

\begin{abstract}
We discuss dynamic spin susceptibility (DSS) in two-dimensional (2D)
Dirac electrons with spin-orbit interactions to characterize topological
insulators.
The imaginary part of the DSS appears as an absorption rate in response
to a transverse ac magnetic field, just as in an electron spin resonance
experiment for localized spin systems.
We found that when the system is in a static magnetic field, the
topological state can be identified by an anomalous resonant peak of the
imaginary part of the DSS as a function of the frequency of the
transverse magnetic field $\omega$.
In the absence of a static magnetic field, the imaginary part of the DSS
becomes a continuous function of $\omega$ with a threshold frequency
$\omega_{\rm c}$. In this case, the topological and the trivial phases
can also be distinguished by the values of $\omega_{\rm c}$ and by the
line shapes.
Thus the DSS is an experimentally observable physical quantity to
characterize a topological insulator directly from bulk properties,
without observing a topological transition.
\end{abstract}

\maketitle 

{\it Introduction.} 
Recently, there is growing interest in the study of two-dimensional (2D)
topological insulators (TIs). This was theoretically predicted by Kane
and Mele~\cite{Kane-M} based on a model describing electrons on a
graphene like honeycomb lattice with spin-orbit interactions.  Although
the first experimental discovery of a TI was in HgTe quantum
wells,~\cite{Konig-2007,Konig-2008,Roth} which is described by the
Bernevig-Hughes-Zhang (BHZ) model,~\cite{Bernevig-H-Z} there are many
candidates of TIs that have a honeycomb lattice structure as was
originally discussed by Kane and Mele, and they are being intensively
studied both theoretically and experimentally. One of these materials is
a silicene,~\cite{Lalmi,Padova-1,Padova-2,Vogt,Lin,Fleurence} which is a
2D crystal of silicon. There are also similar materials called germanene
and stannene, that consist of Ge and Sn,
respectively.~\cite{Liu-F-Y,Liu-J-Y} There are other types of honeycomb
lattice materials that consist of two components, such as molybdenum
dichalcogenides (MoS$_2$, MoSe$_2$, etc.).~\cite{Mak,Xiao} These
materials have buckled honeycomb lattice structures with relevant
intrinsic spin-orbit couplings as compared to graphene.  A tunable band
gap can also be introduced by applying a perpendicular electric field to
the material sheet.

For these systems, the low-energy electronic properties can be described
by the 2D Dirac Hamiltonian, as those of graphenes, with the Fermi
energy at the Dirac point. The band gap and the spin-orbit coupling
appear as the mass term.  In such a Dirac system, there is a topological
phase transition from a TI to a trivial band insulator (BI), at a charge
neutrality point. The TI has quantized spin Hall conductivity when the
spin is conserved, and it is more generally characterized by a $Z_2$
topological number.~\cite{Kane-M} Since the experiment to observe the
topological state depends on transport measurements, it is desirable to
find a non-contact method to identify whether the system is a TI or a BI
from the bulk properties.

There are several physical quantities and experiments proposed to
identify the topological states. For example, optical
responses,~\cite{Ezawa,Stille-T-N,Tabert-N_2013L,Tabert-N_2013B1} spin
and valley Hall effects,~\cite{Dyrdal-B,Tahir-M-S-S,Tabert-N_2013B2}
dynamical polarization
function,~\cite{Tabert-N_2014,Chang-Z-Z-Y,Duppen-V-P} anomalous spin
Nernst effect,~\cite{Gusynin-S-V} quantum oscillations, and orbital
magnetism~\cite{Islam-G,Tsaran-S,Shakouri-V-V-P, Raoux-P-F-M,Tabert-C-N}
have been proposed as such experiments.
However, these experiments are not simple enough to detect TIs, in a
sense that they are not direct information of the topological state.
Especially, in many cases, a TI is identified by observing a topological
transition from a BI.  Therefore, if the internal parameters of the
system are not controllable from the outside, the identification becomes
difficult.

In this Rapid Communication, we turn our attention to the dynamic spin
susceptibility (DSS) whose imaginary part gives us a very simple index
of the topological state only by the existence of a certain resonant
peak structure.  This method enables us to identify a topological phase
directly without observing a topological transition.
We also discuss that an absorption rate in response to a transverse ac
magnetic field is related to the imaginary part of the DSS, just as in
an electron spin resonance (ESR) measurement.

{\it 2D Dirac fermions.}
We consider a Kane-Mele type Hamiltonian~\cite{Kane-M} describing
electrons on a honeycomb lattice with an alternating potential $\Delta$
and a spin-orbit interaction $\kappa$,
\begin{equation}
 \mathcal{H}=t\sum_{\braket{ij}} c_i^\dagger c_j^{\mathstrut}
  + \Delta\sum_i  \eta_i c_i^\dagger c_i^{\mathstrut}
  + \i\frac{\kappa}{3\sqrt{3}} \sum_{\braket{\braket{ij}}}
  \nu_{ij} c_i^\dagger s^z c_j^{\mathstrut},
\end{equation}
where $c_i^{\dag}$ ($c_i$) is a creation (annihilation) operator at site
$i$ (spin indices are omitted).  $\braket{ij}$ and
$\braket{\braket{ij}}$, denote a nearest and a next nearest pair,
respectively. $\eta_i=1$ ($\eta_i=-1$) for the A (B) sublattice, and
$\nu_{ij}=(2/\sqrt{3})(\hat{\bm{d}}_1\times\hat{\bm{d}}_2)_z=\pm 1$,
where $\hat{\bm{d}}_1$ and $\hat{\bm{d}}_2$ are unit vectors along the
two bonds on which the electron hops from a site $j$ to $i$. $s_z$ is a
Pauli matrix describing the electron's spin.

In the continuum limit, the local Hamiltonian in a static magnetic field
is given by
\begin{align}
 \mathcal{H}_{\xi}=&
 \hbar v(\xi\pi_x\tau_x+\pi_y\tau_y)+\Delta\tau_z
 -\kappa\xi\tau_z s_z,
 \label{KM_model}
\end{align}
where $v$ is the Fermi velocity, $\xi=\pm 1$ denotes $K$ and $K'$
points, respectively, $\bm{\pi}=-\i\hbar\nabla+e\bm{A}/c$ is the
momentum operator with $\nabla\times\bm{A}=(0,0,B)$, and $\Delta$ is an
energy gap related to the alternating potential of the A,B sublattices
of the honeycomb lattice.  $\tau_\alpha$ $(\alpha=x,y,z)$ are the Pauli
matrices for the sublattice.

Because the momentum operators $\pi_{\pm}\equiv\pi_x\pm\i\pi_y$ follow
the commutation relation $\left[\pi_+,\pi_-\right]=-\frac{2e\hbar}{c}B$,
they are related to creation and annihilation operators, $a^{\dagger}$
and $a$, as $\pi_{+}=\sqrt{2}\frac{\hbar}{l}a^{\dagger}$,
$\pi_{-}=\sqrt{2}\frac{\hbar}{l}a$, for $eB>0$ with
$l\equiv\sqrt{c\hbar/|e B|}$.  Then the eigenvalues and eigenstates of
the Schr\"{o}dinger equation for valley-$\xi$ and spin-$s$ sector
($s=\pm1$), $\mathcal{H}_{\xi s}|n,\xi,s\rangle\rangle= E_{n}^{\xi
s}|n,\xi,s\rangle\rangle$ are given using the number state $\ket{n}$ of
$a^\dag a^{\mathstrut}$ as
\begin{align}
 E_{n}^{\xi s}&=\left\{
 \begin{array}{ll}
  \sgn(n)
   \sqrt{2\hbar^2v^2|n|/l^2+\Delta_{\xi s}^2}&\quad
   (n\neq 0)\\
  -\xi\Delta_{\xi s}&\quad (n=0)
 \end{array}\right.
  \label{2D_Dirac_eigenvalue},\\
 &
|n,\xi,s\rangle\rangle=
  \left[
 \begin{array}{c}
  A_n^{\xi s}\ket{|n|-\xi_+}\\
  B_n^{\xi s}\ket{|n|-\xi_-}
 \end{array}
 \right],
  \label{2D_Dirac_eigenstate}
\end{align}
where $\Delta_{\xi s}\equiv\Delta-\xi s\kappa$,
$\xi_{\pm}\equiv(1\pm\xi)/2$, and
\begin{subequations} 
\begin{align}
 A_n^{\xi s}
 &=\left\{
 \begin{array}{cc}
  \xi
   \sqrt{|E_n^{\xi s}+\Delta_{\xi s}|/2 |E_n^{\xi s}|}
   &(n\neq 0)\\
  \xi_{-}
   &(n=0)
 \end{array}
 \right.,\\
 B_n^{\xi s}
 &=\left\{
 \begin{array}{cc}
  \sgn(n)
   \sqrt{|E_n^{\xi s}-\Delta_{\xi s}|/2 |E_n^{\xi s}|}
   &(n\neq 0)\\
  \xi_{+}
   &(n=0)
 \end{array}
 \right..
\end{align}
\end{subequations}
Here, the Zeeman effect has been ignored, assuming that the effective
mass of the Dirac system is generally very small, but it can easily be
taken into account only by shifting the Landau levels $\pm
\frac{1}{2}g\mu_{\rm B}B$ according to the directions of the spins.  The
breaking symmetry for positive and negative energies at $n=0$
eigenvalues is called the ``parity anomaly'',~\cite{Haldane} which plays
an essential role in the topological properties of Dirac systems. Due to
the parity anomaly, quantized spin Hall conductivity of the system at
the charge neutrality point with $B=0$ becomes
\begin{equation}
 \sigma_{xy}^s=\frac{e}{2\pi}
  \sgn(\kappa)\theta(|\kappa|-|\Delta|).
\end{equation}
This means that the system is TI for $|\kappa|>|\Delta|$ and BI for
$|\kappa|<|\Delta|$.

\begin{figure}[t]
\begin{center}
 \includegraphics[width=90mm]{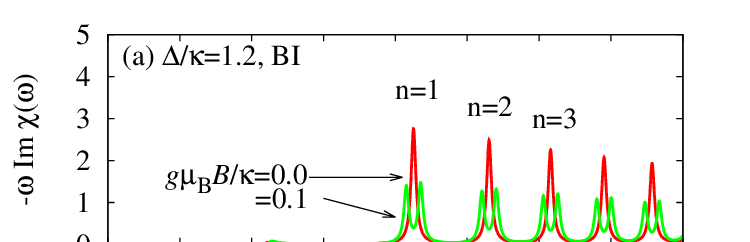}
 \includegraphics[width=90mm]{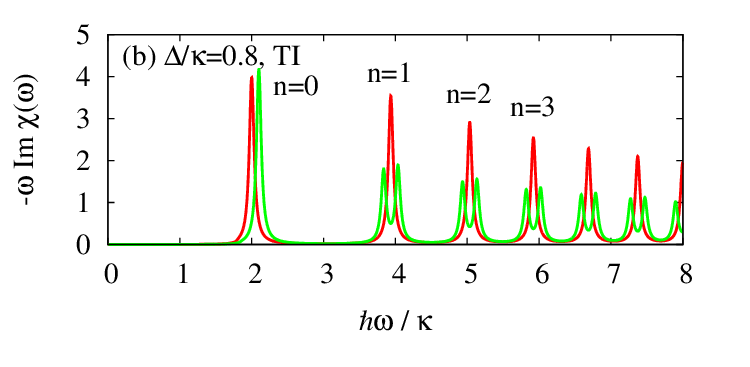}
\end{center}
 \caption{(color online) Imaginary part of the DSS $-\omega\,\Im
 \chi_{xx}(\omega)$ (in units of $\hbar^2/l^2$) with $\mu=0$ and
 $\Gamma/\kappa=0.02$ for (a) BI ($\Delta/\kappa=1.2$) and (b) TI
 ($\Delta/\kappa=0.8$).  The parameters are $B=0.1$ T, $v=5.4\times
 10^5$ m/s, and $\kappa=4.0$ meV assuming a silicene.  There are regular
 peaks for both cases which stem from transitions between $-n$ and $n$
 Landau levels with $|n|\geq 1$. In addition to those, an anomalous peak
 due to $n=0$ Landau levels appears at $\omega=2\kappa/\hbar$
 (independent of the strength of the magnetic field) only for TI. When
 the Zeeman effect exists, e.g., $g\mu_{\rm B}B/\kappa=0.1$, the regular
 peaks split into two, while the anomalous peak shifts only to one
 direction.}
\label{fig1}
\end{figure}

\begin{figure}[t]
\begin{center}
 \includegraphics[width=70mm]{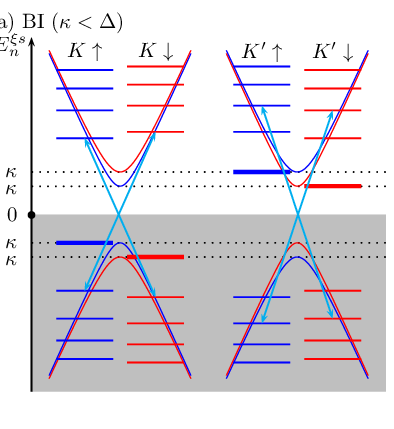}
 \includegraphics[width=70mm]{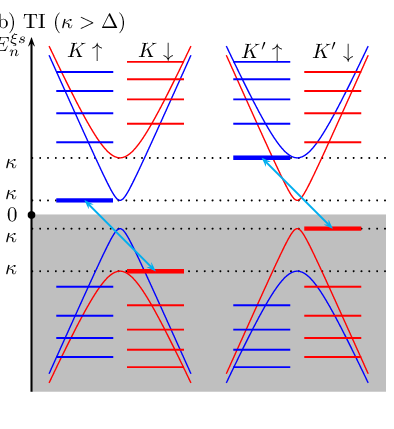}
\end{center}
 \caption{(color online) Landau level structures of Dirac fermions with
 alternating potential $\Delta$ and spin-orbit interactions $\kappa$ for
 (a) the BI ($|\kappa|<|\Delta|$) and (b) the TI ($|\kappa|>|\Delta|$)
 states. The two phases are characterized by the configurations of the
 $n=0$ Landau levels around the Fermi level ($\mu=0$). For the BI, the
 $n=0$ Landau levels have an opposite sign in each spin, so that the
 quantized Hall conductivity canceled out when the magnetic field is
 tuned off, and the absorption rate does not show the anomalous peak,
 because the transitions between $n=0$ Landau levels are not allowed.}
\label{fig2}
\end{figure}

{\it Properties of DSS.}
Now let us consider the properties of DSS in these two phases of the
Dirac system.  The spin-spin correlation function is given in a
Matsubara form as
\begin{align}
 \chi_{xx}(\tau)=&-\braket{\mathcal{T}_{\tau}S^x(\tau)S^x(0)}_0,\\
 \chi_{xx}(\i\nu_m)=&\frac{1}{\beta\hbar}
 \sum_{k,n}
 \Tr
 \mathcal{G}(\varepsilon_k,\i\omega_n)(\tau_0 s_x)
 \mathcal{G}(\varepsilon_k,\i\omega_{n}^+)(\tau_0 s_x),
 \label{eqn:Matsubara_sum1}
\end{align}
where $\Braket{\cdots}_0\equiv
\Tr\cdots\e^{-\beta\mathcal{H}}/\Tr\e^{-\beta\mathcal{H}}$, with inverse
temperature $\beta$.  $\tau_0$ is the unit matrix, and the temperature
Green's function is given as $\mathcal{G}(\varepsilon,\i\omega_n)
=-(\i\omega_n-(\varepsilon-\mu)/\hbar+\i\,\sgn(\omega_n)\Gamma/\hbar)^{-1}$,
$\omega_{n}^+\equiv \omega_n+\nu_m$ with $\omega_{n}$ and $\nu_m$ being
Matsubara frequencies for fermions and bosons, respectively.  The
chemical potential and the impurity scattering time are denoted by
$\mu(=0)$ and $\hbar/2\Gamma$, respectively.  After analytic
continuation $\i\nu_m\to\omega+\i 0$, the retarded spin-spin correlation
function is obtained as
\begin{align}
&
\chi_{xx}(\omega)=
  \frac{\hbar^2}{8\pi l^2}
 \hbar\int_{-\infty}^{\infty}\d
 \Omega[-f'(\hbar\Omega)]
 \sum_{\xi,s}\sum_{n,m}
\nonumber\\
&\times
 \mathcal{X}(
 E_{n}^{\xi s},
 E_{m}^{\xi\bar{s}},\omega;\Omega)
 (A_n^{\xi s}A_m^{\xi\bar{s}}
 +B_n^{\xi s}B_m^{\xi\bar{s}})^{2}
 \delta_{|n|,|m|},
\label{chi.g1}
\end{align}
where $f(\varepsilon)=(\e^{\beta (\varepsilon-\mu)}+1)^{-1}$ is the
Fermi distribution function, $\bar{s}$ means the opposite spin of
$s$, and $\mathcal{X}(x,y,\omega;\Omega)$ is defined as follows
\begin{align}
\lefteqn{
 \left.\frac{1}{\beta\hbar}\sum_n
 \frac{1}{(\i\omega_n-x/\hbar)(\i\omega_n^+-y/\hbar)}
 \right|_{\i\nu_m\to\omega}}\nonumber\\
&
=\hbar\int_{-\infty}^{\infty}\d
 \Omega[-f'(\hbar\Omega)]\mathcal{X}(x,y,\omega;\Omega).
\end{align}
The transverse spin susceptibility is given by energy transitions
between Landau levels labeled by the same absolute value of $n$ with
opposite signs. This selection rule is much simpler than that in the
current-current correlation function for the optical conductivity, which
is related to the transitions between the $|n|$ and $|n\pm 1|$ Landau
levels.~\cite{Stille-T-N,Tabert-N_2013L,Tabert-N_2013B1,Tabert-N_2013B2}

Now we turn our attention only to the imaginary part of the DSS
multiplied by the frequency $-\omega\,\Im \chi_{xx}(\omega)$, which is
experimentally observable as discussed later.  The real part can also be
obtained via the Kramers-Kronig relation.  As shown in Fig.~\ref{fig1},
$-\omega\,\Im \chi_{xx}(\omega)$ at $\mu=0$ has several peaks
corresponding to the transitions between the Landau levels below and
above the Fermi level with the same absolute values of the index $|n|\ge
1$. We also find that an anomalous peak appears only for the TI region
($|\kappa|>|\Delta|$), {\it whose resonant frequency is independent of
the strength of the magnetic field}.  Actually, for the clean and zero
temperature limit ($\Gamma\to 0$ and $T \to 0$), we get the following
$\delta$-function peaks,
\begin{align}
&
 -\Im \chi_{xx}(\omega)=
 \frac{\hbar^2}{4 l^2}
 \biggl[
 \sum_{n=1}^{\infty}
 2(A_n^{+1,\uparrow}B_{n}^{+1,\downarrow}
 -B_n^{+1,\uparrow}A_{n}^{+1,\downarrow})^{2}
\nonumber\\
 &
\times
 \delta(\omega-(E_{n}^{+1,\uparrow}+E_{n}^{+1,\downarrow})/\hbar)
 +
\theta(\kappa^2-\Delta^2)\delta(\omega-2|\kappa|/\hbar)
\biggr].
 \label{a-rate.g0}
\end{align}
The last term of Eq.~(\ref{a-rate.g0}) is the anomalous peak at
$\omega=2\kappa/\hbar$ which stems from the transition between the $n=0$
Landau levels with opposite spins.  This peak can be observed only in
the TI where one of the two levels is below the Fermi level, as shown in
Fig.~\ref{fig2}.
Physically, this peak is related to an energy $2\kappa$ to make an edge
state with the spin current in the opposite direction via spin flipping.
Therefore, we can identify whether the system is a TI or a BI, by
observing this anomalous peak.  The numerical value of the frequency of
the anomalous peaks is estimated as $\omega\approx 12$ THz for silicene
with $\kappa=4.0$ meV.~\cite{Tsai}

\begin{figure}[t]
\begin{center}
 \includegraphics[width=90mm]{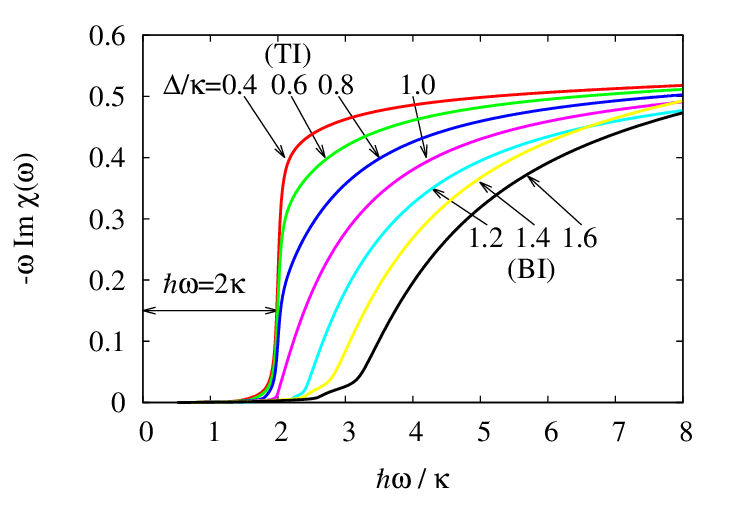}
\end{center}
 \caption{(color online) Imaginary part of the DSS,
 $-\omega\,\Im\chi_{xx}(\omega)$ (in unit of $\kappa^2/v^2$), with $B=0$
 and $\Gamma/\kappa=0.02$ for several values of
 $\Delta/\kappa=0.4$-$1.6$.  For small $\Gamma$,
 $-\omega\,\Im\chi_{xx}(\omega)$ is discontinuous (continuous) function
 at the threshold frequency $\omega_{\rm c}=2\kappa/\hbar$ ($\omega_{\rm
 c}=2\Delta/\hbar$) for TI (BI).}
\label{fig3}
\end{figure}

We can also show that the anomalous peak can easily be distinguished
from the regular peaks, even if the Zeeman effect is taken into account.
In the presence of the Zeeman effect, Eq.~(\ref{chi.g1}) is modified as
$\mathcal{X}( E_{n}^{\xi s}, E_{m}^{\xi\bar{s}},\omega;\Omega)$ $\to$
$\mathcal{X}( E_{n}^{\xi s}+s g\mu_{\rm B}B/2,
E_{m}^{\xi\bar{s}}+\bar{s} g\mu_{\rm B}B/2,\omega;\Omega)$.  As shown in
Fig.~\ref{fig1}, the regular peaks split into two parts
$\Delta\omega=\pm\mu_{\rm B}B/\hbar$, while the anomalous peak shifts to
one direction $\Delta\omega=\sgn(\kappa)\mu_{\rm B}B/\hbar$.  This is
because the valley degeneracy is lifted by the Zeeman effect for
$|n|\geq 1$ Landau levels, while the Landau levels are uniformly shifted
for $n=0$.  Therefore, the DSS is still effective information to
characterize the TI in the presence of the Zeeman coupling, if it is
sufficiently smaller than the spin-orbit coupling, as usually expected
for TIs. Furthermore the Zeeman shift of the anomalous peak has
information to identify the sign of the spin-orbit coupling $\kappa$.

{\it Case without static magnetic field.}
So far we have assumed the Landau quantization by a static magnetic
field in a bulk system, so that the peak structures of $\Im
\chi_{xx}(\omega)$ vanish in the $B=0$ cases, and it becomes a
continuous function of $\omega$. However, as shown in Fig.~\ref{fig3},
the threshold value of the frequency $\omega_{\rm c}$ which gives the
minimum edge of finite $-\omega\,\Im\chi_{xx}(\omega)$ is $\omega_{\rm
c}=2\Delta/\hbar$ for the BI and $\omega_{\rm c}=2\kappa/\hbar$ for the
TI. This difference can be easily understood from Eq.~(\ref{a-rate.g0})
for $B\to 0$ which gives $\hbar\omega_{\rm
c}=|\Delta_{+1,\uparrow}|+|\Delta_{+1,\downarrow}|$.  The line shape of
$\Im \chi_{xx}(\omega)$ is also different for the two phases: $\Im
\chi_{xx}(\omega)$ grows continuously from $0$ for the BI while it grows
discontinuously for the TI.  This can also be understood by examining
the $B\to 0$ limit of Eq.~(\ref{a-rate.g0}): For $|\Delta|\gg|\kappa|$,
we get $-\Im \chi_{xx}(2\Delta/\hbar)=0$ and
$\lim_{\omega\to\infty}[-\omega\:\Im \chi_{xx}(\omega)]
=\frac{\kappa^2}{4v^2}$, while for $|\Delta|\ll|\kappa|$, we have $-\Im
\chi_{xx}(2\kappa/\hbar)=\frac{\hbar\kappa}{4v^2}$ and
$\lim_{\omega\to\infty}[-\omega\:\Im \chi_{xx}(\omega)]
=\frac{2\kappa^2-\Delta^2}{4v^2}$.  These features may also give
supplemental information to distinguish a TI and a BI.

{\it Possible experiments.}
Finally, let us consider how we observe $\Im \chi_{xx}(\omega)$ in real
experiments. For an example, we examine a 2D electron system with a
static magnetic field and a transverse ac magnetic field.  The
Hamiltonian of the system is expressed as
\begin{equation}
 \mathcal{H}(t)=\mathcal{H}+V(t),\label{Ham}
\end{equation}
where $\mathcal{H}$ and $V(t)$ are the time-independent and
time-dependent parts of the Hamiltonian, respectively.  For the static
and the dynamic magnetic fields,
\begin{equation}
 \bm{B}(t)=\bm{B}+\bm{B}_{\rm R}\cos(\omega t),
\end{equation}
with $\bm{B}=(0,0,B)$ and $\bm{B}_{\rm R}=(B_{\rm R},0,0)$, we may
choose the vector potential as $\bm{A}=B(-y/2,x/2,0)$, $\bm{A}_{\rm
R}=B_{\rm R}(0,0,y)$.  In order to discuss general situations, we first
consider a ``non-relativistic'' system with a parabolic band, and with
an electron mass $m$. Then Eq.~(\ref{Ham}) is given by
\begin{align}
 \mathcal{H}=&\mathcal{H}_0(\bm{A})
 +g\mu_{\rm B}B\int\d\bm{r}S^z(\bm{r}),\\
 V(t)=&
 g\mu_{\rm B}B_{\rm R}\cos(\omega t)
 \int\d\bm{r}S^x(\bm{r})\nonumber\\
 &-\frac{\cos^2(\omega t)}{2c}\int\d\bm{r}
 \bm{A}_{\rm R}(\bm{r})\cdot\bm{J}_{\rm d}(\bm{r}),
 \label{Ham_Vt}
\end{align}
where $S^\alpha\equiv\int \d\bm{r}\hat{\psi}_{s}^{\dag}
s^{\alpha}_{ss'}\hat{\psi}_{s'}^{\mathstrut}$ with an electron operator
$\hat{\psi}_{s}$.  $\bm{J}_{\rm d}\equiv -\frac{e^2}{mc}\bm{A}_{\rm R}
\hat{\psi}_{s}^{\dagger} \hat{\psi}_{s}^{\mathstrut}$ means a
diamagnetic current induced by the dynamic magnetic field, whereas a
paramagnetic current $\bm{J}_{\rm p}\equiv -\frac{e\hbar}{2m\i} \{
\hat{\psi}^{\dagger}_{s} \nabla\hat{\psi}_{s}^{\mathstrut}
-[\nabla\hat{\psi}^{\dagger}_{s}(\bm{r})] \hat{\psi}^{\mathstrut}_{s}
\}$ is included in $\mathcal{H}_0$.  We have also used the relation
$\bm{A}_{\rm R}(\bm{r})\cdot\bm{J}_{\rm p}(\bm{r})=0$, which is
satisfied in 2D systems.

Then the absorption rate of the dynamic magnetic field is obtained up to
the second order of $V$ as
\begin{align}
 I(\omega)&=
 \int_{0}^{T}
 \frac{\d t'}{T}\frac{\d}{\d t'}\Tr[\rho(t')\mathcal{H}(t')]
\\
&
 \approx 
 -\frac{(g\mu_{\rm B}B_{\rm R})^2\omega}{2\hbar}
 \Im\chi_{xx}^{\rm R}(\omega),
 \label{ESRAR}
\end{align}
where $T\equiv2\pi/\omega$, $\rho$ is a density matrix of
Eq.~(\ref{Ham}), and $\chi_{xx}^{\rm R}(\omega)$ is the Fourier
transform of the retarded transverse spin-spin correlation function,
\begin{equation}
 \chi_{xx}^{\rm R}(t)=-\i\theta(t)
  \Braket{\left[S_{\rm tot}^x(t),S_{\rm tot}^x(0)\right]}_0.
\end{equation}
%
%
Here, we should note that the orbital contributions of the ac field do
not appear in Eq.~(\ref{ESRAR}), because $\bm{A}_{\rm
R}(\bm{r})\cdot\bm{J}_{\rm d}(\bm{r})$ is canceled in the time integral,
and $I(\omega)$ becomes just the same form as a usual ESR formula for
localized spin systems.~\cite{Kubo-T} This situation is realized only in
2D systems where the electrons do not have a path to move along the $z$
direction.  In Dirac systems, the current operator does not have a
diamagnetic part because of the linear dispersion relation, so that the
last term of the right hand side of Eq.~(\ref{Ham_Vt}) does not
exist. Thus Eq.~(\ref{ESRAR}) can also be applied to current 2D Dirac
systems.

{\it Summary.}
We have discussed the dynamic transverse spin susceptibility (DSS) of a
2D Dirac system with a spin-orbit interaction $\kappa$ and an
alternating potential $\Delta$ (\ref{KM_model}). Its imaginary part is
related to the absorption rate in response to a transverse ac magnetic
field.  When a static magnetic field is applied to the system, the
imaginary part of the DSS shows an anomalous peak at frequency
$\omega=2\kappa/\hbar$ for the TI.
This is related to an energy to make an edge state with the spin current
in the opposite direction via spin flipping.
On the other hand, when the static magnetic field is turned off, the
imaginary part of the DSS becomes a continuous function of $\omega$ with
different values of threshold frequencies, $\omega_{\rm
c}=2\kappa/\hbar$ for the TI and $\omega_{\rm c}=2\Delta/\hbar$ for the
BI, respectively.
These properties enable us to identify the TI directly only from the
bulk information, without observing a topological transition from the
BI.
The DSS is considered to be a relevant probe for TIs in more general
cases. However, in the case of the BHZ model\cite{Bernevig-H-Z} which
describes a TI realized in HgTe quantum wells, the present theory cannot
be applied straightforwardly, since the $n=0$ Landau levels depend on
the strength of the magnetic field\cite{Konig-2007}. We also need
extended discussions for the original Kane-Mele model including the
Rashba interactions\cite{Kane-M} where the $z$-component of spins is
not conserved.

%
The authors thank S.~C.~Furuya, Y. Kubo, and N.~Hatano for helpful
discussions.  M.~N. thanks the Max Planck Institute for Solid State
Research where a part of this work has been done.

\end{document}